\documentclass[twocolumn,showpacs,preprintnumbers,amsmath,amssymb]{revtex4}

\usepackage{graphicx}
\usepackage{dcolumn}
\usepackage{amsmath}
\usepackage{amssymb}

\def\vk{{\bf k}}

\def\vS{{\bf S}}

\def\bra{\langle}
\def\ket{\rangle}

\newcommand{\eq}[1]{Eq.~(\ref{#1})}
\newcommand{\fig}[1]{Fig.~\ref{#1}}
\newcommand{\be}{\begin{equation}}
\newcommand{\ee}{\end{equation}}
\newcommand{\bea}{\begin{eqnarray}}
\newcommand{\no}{\nonumber}
\newcommand{\eea}{\end{eqnarray}}
\newcommand{\bean}{\begin{eqnarray*}}
\newcommand{\eean}{\end{eqnarray*}}
\newcommand{\bfi}{\begin{figure}}
\newcommand{\efi}{\end{figure}}
\newcommand{\bc}{\begin{center}}
\newcommand{\ec}{\end{center}}
\newcommand{\ba}{\begin{array}}
\newcommand{\ea}{\end{array}}

\begin{document}


\title{Self-masking of spontaneous symmetry breaking in layer materials}

\author{Hiroyuki Yamase} 
\affiliation{Max-Planck-Institute for Solid State Research, 
Heisenbergstrasse 1, D-70569 Stuttgart, Germany}

\date{\today}

\begin{abstract} 
We study $d$-wave Fermi surface deformations ($d$FSD), the so-called 
Pomeranchuk instability, 
on bilayer and infinite-layer square lattices. 
Since the order parameter of the $d$FSD has 
Ising symmetry, 
there are two stacking patterns along the $c$ axis: 
$(+,+)$ and $(+,-)$. 
We find that, as long as the $c$ axis dispersion is finite 
at the saddle points of the in-plane band dispersion, 
the $(+,-)$ stacking is usually favored independently 
on the details of interlayer coupling, yielding no macroscopic anisotropy. 
The $d$FSD provides unique spontaneous symmetry breaking that 
is self-masked in layer materials.  
\end{abstract}

\maketitle

Spontaneous symmetry breaking is a key concept in physics. 
Typical examples in condensed matter are crystallization, 
ferromagnetism, antiferromagnetism, ferroelectricity, 
and superconductivity. 
Recently a new type of symmetry breaking 
was found in the shape of the Fermi surface (FS) \cite{yamase00,metzner00}. 
The FS usually fulfills the point-group symmetry of the 
underlying lattice structure. 
However, it was shown that such symmetry of the FS 
can be broken by electron correlation effects in the 
two-dimensional (2D) $t$-$J$ \cite{yamase00,miyanaga06,edegger06} and 
Hubbard \cite{metzner00,valenzuela01,hankevych02,neumayr03} 
models on a square lattice: 
The FS expands along the $k_x$ direction 
and shrinks along the $k_y$ direction, or vice versa. 
This instability is characterized by a $d$-wave order parameter  
and we refer to it as   
$d$-wave Fermi surface deformations ($d$FSD) or $d$-wave Pomeranchuk 
instability. 

The $d$FSD state has the same 
symmetry as the electronic nematic state \cite{kivelson98}.   
Implications of such reduced symmetry were 
widely discussed for high-$T_{c}$ cuprates \cite{kivelson03}.  
In particular, the strong $ab$-anisotropy of magnetic 
excitation spectrum observed for untwinned 
YBa$_{2}$Cu$_{3}$O$_{y}$ \cite{hinkov0407} was well explained 
in terms of the $d$FSD and its competition with singlet pairing \cite{yamase06}. 
Moreover many peculiar features observed for the 
bilayer ruthenate  Sr$_{3}$Ru$_{2}$O$_{7}$  \cite{grigera04,borzi07} 
also turned out to be well captured in terms of 
the $d$FSD instability \cite{kee05,doh07,yamase07bc,ho08}.

The $d$FSD is expected in layer materials, which are characterized by 
weak interlayer coupling. 
Since the order parameter of the $d$FSD has Ising symmetry,  
the interlayer coupling drives two possible 
stacking patterns between the layers, $(+,+)$ and $(+,-)$;  
we call the former {\it ferro-type} (F) stacking and 
the latter {\it antiferro-type} (AF). 
In the latter case, 
macroscopic anisotropy does not appear, 
leading to self-masking of the underlying $d$FSD instability.

In this Letter, we propose a microscopic theory for the $d$FSD phenomenon 
in realistic layer materials. 
We find that AF $d$FSD state is usually favored independently on 
the details of interlayer coupling. 
Thus the $d$FSD instability provides unique spontaneous symmetry breaking 
that is self-masked macroscopically. 
This generic conclusion 
is applicable to a wide range of layer materials, 
including recently discovered iron-based superconductors 
\cite{fang08,xu08}.

We consider both bilayer and infinite-layer models. 
We first focus on a bilayer square lattice and 
analyze the following pure forward scattering model, 
\bea
&&H = \sum_{\substack{\vk,\, \sigma \\ i=A,B}} 
(\epsilon_{\vk} - \mu) 
c^{i\,\dagger}_{\vk\sigma} c^{i}_{\vk\sigma} + 
 \frac{1}{2N} \sum_{\substack{\vk,\, \vk' \\ i=A,B}} 
f_{\vk\vk'} \, n^{i}_{\vk} n^{i}_{\vk'}  \no \\
&&\qquad\qquad\qquad +\sum_{\vk,\,\sigma} \epsilon_{\vk}^{z} (
c^{A\,\dagger}_{\vk \sigma} c^{B}_{\vk \sigma} + 
c^{B\,\dagger}_{\vk \sigma} c^{A}_{\vk \sigma}
)\,, \label{model}
\eea 
where $c^{i\,\dagger}_{\vk\sigma} (c^{i}_{\vk\sigma})$ 
creates (annihilates) an electron with momentum $\vk$ and 
spin $\sigma$ on the $i=A$ and $B$ planes; 
$n^{i}_{\vk} = \sum_{\sigma}c^{i\,\dagger}_{\vk\sigma} 
c^{i}_{\vk\sigma}$ is the number operator; 
$N$ is the total number of sites on the $i$ plane and 
$\mu$ is the chemical potential. 
For hopping amplitudes $t$ and $t'$ between nearest and next-nearest 
neighbors on the square lattice, $\epsilon_{\vk}$ is given by 
$\epsilon_{\vk} = -2 t (\cos k_{x}+\cos k_{y})- 4t'\cos k_{x} \cos k_{y}$. 
The forward scattering interaction, $f_{\vk \vk'}=-g d_{\vk} d_{\vk'}$, 
drives the $d$FSD instability, where $d_{\vk}=\cos k_{x} -\cos k_{y}$ 
and $g>0$ \cite{khavkine04,yamase05}.   
The last term in Hamiltonian (\ref{model}) is the hybridization between 
A and B planes. 
We consider three types of $c$ axis dispersions, 
$\epsilon_{\vk}^{z}=-t_{z}$, 
$-2t_{z} (\cos k_x + \cos k_y)$, and  
$-4t_{z} \cos\frac{k_x}{2} \cos\frac{k_y}{2}$; 
the latter two dispersions vanish at $\vk=(\pi,0)$ and $(0,\pi)$ while 
the former does not. This difference is crucial to the stacking 
of the $d$FSD.

Hamiltonian (\ref{model}) is analyzed in the Hartree approximation, 
which becomes exact in the thermodynamic limit in our model. 
We obtain the mean field 
\be
\eta^{A (B)}=-\frac{g}{N} \sum_{\vk} d_{\vk} \bra n_{\vk}^{A (B)} \ket \,, 
\ee
which is nonzero only if the electronic state loses the fourfold 
symmetry of the square lattice and is thus the order parameter of the 
$d$FSD in the $A (B)$ plane. 
The FS is elongated along the $k_{x}$ and $k_{y}$ directions 
for $\eta^{A(B)}>0$ and $\eta^{A(B)}<0$, respectively, 
i.e., the order parameter has Ising symmetry. AF (F) stacking 
is defined by $\eta^{A} \eta^{B}<0 (>0)$.  
We determine the mean fields self-consistently under 
the constraint that each plane has the same electron 
density. 
We choose $t'/t=0.35$ \cite{yamase07bc,liu08} and 
$t_{z}/t=0.1$ \cite{liu08}, 
which is appropriate for the bilayer ruthenate. 
$\epsilon_{\vk}$  has saddle points in 
$\vk=(\pi,0)$ and $(0,\pi)$, and thus 
the van Hove energy of $\epsilon_{\vk}$ 
is given by $\mu_{\rm vH}^{0}=4t'=1.4t$. 
In the following we set $t=1$ and shift the energy 
such that $\mu_{\rm vH}^{0}\equiv 0$.

\begin{figure*}[ht]
\centerline{\includegraphics[width=0.68\textwidth]{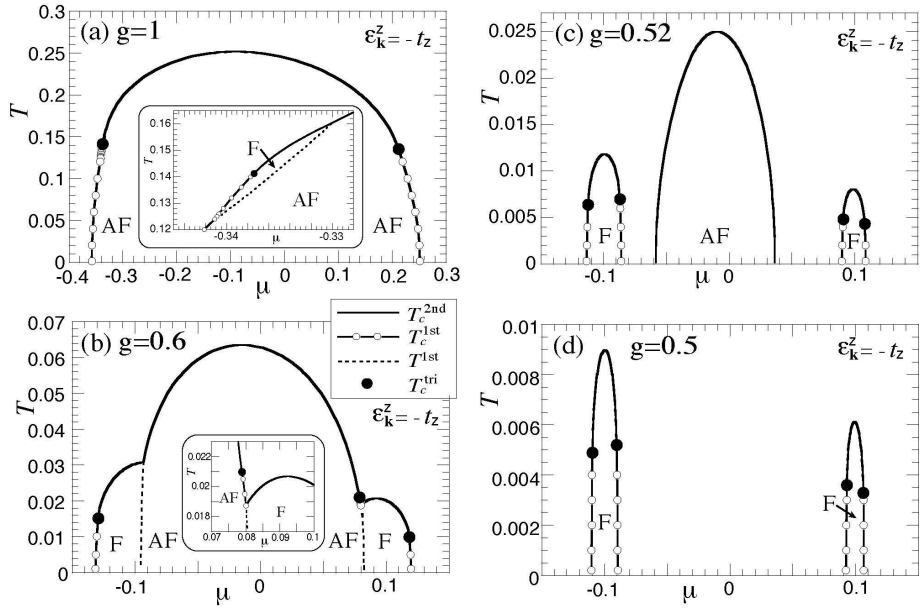}}
\caption{$\mu$-$T$ phase diagrams in the bilayer model with 
$\epsilon_{\vk}^{z}=-t_{z}$ for several choices of $g$ 
at $t_{z}=0.1$. 
Solid lines, $T_{c}^{\rm 2nd}$, denote second order transitions, 
while first order transitions are denoted by open circles, 
$T_{c}^{\rm 1st}$, and dotted lines, $T^{\rm 1st}$; 
the latter, present in panels (a) and (b), 
corresponds to a transition between F and AF;  
solid circles represent tricritical points. 
The insets 
magnify the regions around $\mu\approx -0.33$ and $T\approx 0.15$ 
in (a), and around $\mu\approx 0.08$ and $T\approx 0.02$ in (b).}
\label{z0phase}
\end{figure*}

We first consider the $c$ axis dispersion, $\epsilon_{\vk}^{z}=-t_{z}$, 
from which we can extract 
a generic conclusion about the stacking of the $d$FSD.  
Figure~\ref{z0phase}(a) shows the phase diagram obtained for $g=1$. 
The $d$FSD instability occurs around the van Hove energy 
of the in-plane dispersion, i.e. $\mu$=0 
as in the case of the single-layer model (Ref.~\onlinecite{yamase05}); 
$T_{c}$ is almost unchanged  
by the presence of $\epsilon_{\vk}^{z}$. 
The AF $d$FSD state, namely $\eta^{A} = - \eta^{B} \ne 0$, 
is stable in most of the region of the phase diagram. 
The F stacking appears in a very small region,  
as shown in the inset of \fig{z0phase}(a). 
When the coupling constant $g$ is reduced, the $F$ regions increase, 
close to the edges of the transition line. 
Yet the AF region is still much larger than the F ones, 
as it is shown in \fig{z0phase}(b). 
Upon further reducing $g$, the F and AF regions split from 
one another [\fig{z0phase}(c)]. 
After this \lq\lq intermediate'' phase, the property of the phase diagram 
drastically changes  
below $g = 0.5$, as shown in \fig{z0phase}(d):  
No $d$FSD instability appears around $\mu$=0, but the transition 
occurs around the van Hove energy of the bonding and antibonding bands, 
namely around $\mu_{\rm vH}=\pm t_{z}= \pm 0.1$.  
The phase diagram eventually contains only F regions. 
This property does not change down to an infinitesimally 
small $g$.

To gain an understanding of the $d$FSD stacking, 
we consider the Landau expansion of the grand canonical 
potential (per lattice site) 
with respect to $\eta^{A}$ and $\eta^{B}$, 
\be
\omega(\eta^{A},\eta^{B})=\frac{\mathcal{I}}{2}\left[
(\eta^{A})^{2} + (\eta^{B})^{2}\right] + 
\mathcal{J} \eta^{A}\eta^{B} +\cdots \label{Landau}\,. 
\ee
The condition, $\mathcal{I}^{2}-\mathcal{J}^{2}=0$, 
determines a second order transition. 
$\mathcal{J}$ drives AF and F stacking for $\mathcal{J}>0$ and 
$\mathcal{J}<0$, respectively, and is given by 
\be
\mathcal{J}=\frac{1}{2N}\sum_{\vk} d_{\vk}^{2}
\left\{
f'(\lambda_{\vk}^{+})+f'(\lambda_{\vk}^{-})-
\frac{f(\lambda_{\vk}^{+})-f(\lambda_{\vk}^{-})}{\epsilon_{\vk}^{z}}
\right\}\,, \label{Landau-c}
\ee
where $\lambda_{\vk}^{\pm}=\epsilon_{\vk}-\mu\pm \epsilon_{\vk}^{z}$ 
is the dispersion of the bonding and antibonding bands; 
$f(\lambda)=({\rm e}^{\lambda/T}+1)^{-1}$ is the Fermi function 
at temperature $T$ and $f'$ is its derivative. 

As long as a second order transition occurs around $\mu$=0  
[Figs.~\ref{z0phase}(a) and (b)], we can analyze \eq{Landau-c} 
by expanding it with respect to 
$\epsilon_{\vk}^{z}$ and obtain 
\be
\mathcal{J}=\frac{1}{3}\frac{\partial^{2}}{\partial \mu^{2}}
\int{\rm d}\epsilon N_{d}^{z}(\epsilon)f'(\epsilon-\mu)
\label{Landau-c3}
\ee
where 
\be
N_{d}^{z}(\epsilon)=\frac{1}{N}\sum_{\vk}d_{\vk}^{2}
(\epsilon_{\vk}^{z})^{2} \delta(\epsilon_{\vk}-\epsilon)\,.
\label{dzDOS}
\ee
Since the $\vk$ summation in \eq{dzDOS} is dominated by the contribution 
from the saddle points of $\epsilon_{\vk}$, we 
expand $\epsilon_{\vk}$ and $\epsilon_{\vk}^{z}$ around the saddle points. 
$\epsilon_{\vk}$ is then written as 
$\epsilon_{\vk}=\frac{k_{+}k_{-}}{2m}$ after choosing a suitable choice of 
momentum variables $k_{+}$ and $k_{-}$ with a cutoff momentum $\Lambda$, 
namely $|k_{\pm}| < \Lambda$; 
$m$ is related to $t$ and $t'$; 
$d_{\vk}$ is replaced by a constant, which just 
rescales $\epsilon_{\vk}^{z}$, and we set $d_{\vk}=1$.

\begin{figure}
\centerline{\includegraphics[width=0.45\textwidth]{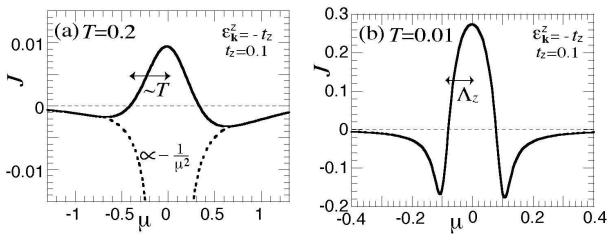}}
\caption{Coupling constant $\mathcal{J}$ between $\eta^{A}$ and $\eta^{B}$ 
as a function of $\mu$ at $T=0.2$ (a) and $0.01$ (b) for 
$\epsilon_{\vk}^{z}=-t_{z}$. In (a), the function of $-\mu^{-2}$ 
is also drawn.} 
\label{c}
\end{figure}

For $\epsilon_{\vk}^{z}=-t_{z}$, 
\eq{dzDOS} becomes 
\be
N_{d}^{z}(\epsilon)=\frac{2mt_{z}^{2}}{\pi^{2}}\log 
\frac{\epsilon_{\Lambda}}{|\epsilon|} 
\label{dzDOS-2}
\ee
where $\epsilon_{\Lambda}=\frac{\Lambda^{2}}{2m}$. 
At zero temperature \eq{Landau-c3} reads 
$\mathcal{J}=-\frac{2mt_{z}^{2}}{3\pi^{2}}\frac{1}{\mu^{2}}$.  
This $\mu^{-2}$ singularity is, however,  
cut off at a scale of the order of $T_{c}^{0}$, 
the transition temperature at $\mu=0$. 
Therefore $\mathcal{J}$ becomes positive 
for $|\mu| \lesssim T_{c}^{0}$, as shown in \fig{c}(a). 
Because of the universal property of the $d$FSD instability 
obtained for a single layer model \cite{yamase05},  
the $d$FSD phase is stabilized in the region $|\mu| \lesssim T_{c}^{0}$. 
Therefore the sign of $\mathcal{J}$ is expected to be positive, 
leading to AF stacking of the $d$FSD in a major region of 
the phase diagram, as seen in Figs.~\ref{z0phase} (a) and (b).

If the coupling constant $g$ is reduced, 
the temperature scale of the $d$FSD becomes small and 
the energy scale of $\epsilon_{\vk}^{z}$ must be taken into account. 
We consider \eq{Landau-c} 
instead of its lowest order expansion of \eq{Landau-c3}. 
The scale of $\epsilon_{\vk}^{z}$ is defined as the bilayer splitting 
at the saddle points of the in-plane dispersion, namely 
\be
\Lambda_{z}=|\epsilon_{\vk}^{z}|\; {\rm at} \;
\vk=(\pi,0) \; {\rm and} \; (0,\pi)\, 
\ee 
in the present case and thus $\Lambda_{z}=t_{z}$. 
The positive $\mathcal{J}$ region is bounded by the scale 
$\Lambda_{z}$, rather than $T$, as seen in \fig{c}(b). 
Then $\mathcal{J}$ becomes negative for  $|\mu| \gtrsim \Lambda_{z}$  
and shows steep dips at
the van Hove energy of the bonding and 
antibonding bands, i.e. at $\mu_{\rm vH}=\pm t_{z}$. 
This structure originates from the 
singular contribution of $f'$ in \eq{Landau-c}. 
When the temperature scale becomes much smaller than $\Lambda_{z}$,  
we have seen that the $d$FSD instability occurs around $\mu=\mu_{\rm vH}$. 
Therefore we obtain F stacking, as seen in \fig{z0phase}(d).

So far we have considered a simple $c$ axis dispersion.  
However, the above analysis 
holds also for other choices of a $c$ axis dispersion, e.g.  
$\epsilon_{\vk}^{z}=-t_{z} (\cos k_x - \cos k_y)^{2}$, 
as long as $\Lambda_{z}\ne 0$. 
Hence we obtain a generic conclusion for bilayer systems: 
F stacking prevails when the temperature scale  
of the $d$FSD is much smaller than $\Lambda_{z}$, otherwise 
the major stacking is AF \cite{misc-kzdispersion}. 

What happens for $\Lambda_{z}=0$? This can be seen from 
the analysis of \eq{Landau-c3}. When $\epsilon_{\vk}^{z}$ is expanded as 
$\epsilon_{\vk}^{z}=(a_{1}k_{+}+b_{1}k_{-})t_{z}$ around the saddle points 
of the in-plane band dispersion, the logarithmic singularity in \eq{dzDOS-2} 
is weakened to $\epsilon \log |\epsilon|$, leading to 
$\mathcal{J}\approx \frac{8m^{2}t_{z}^{2}}{3\pi^{2}}\frac{a_{1}b_{1}}{\mu}$. 
The dispersion in this case may be taken as 
$\epsilon_{\vk}^{z}=-4t_{z} \cos\frac{k_x}{2} \cos\frac{k_y}{2}$, 
for which we obtain $a_{1}b_{1}<0$.  
The phase diagram indeed showed AF and F stacking for $\mu<0$ and $\mu>0$. 
On the other hand, when $\epsilon_{\vk}^{z}$ is expanded as 
$\epsilon_{\vk}^{z}=(a_{2}k_{+}^{2}+b_{2}k_{-}^{2}+c_{2}k_{+}k_{-})t_{z}$, we obtain 
$\mathcal{J}\approx \frac{16m^{3}t_{z}^{2}}{3\pi^{2}}(2a_{2}b_{2}+c_{2}^{2})\log\frac{|\mu|}{\epsilon_{\Lambda}}$. 
Thus F (AF) stacking is stabilized 
for $2a_{2}b_{2}+c_{2}^{2}>0 (<0)$; 
for $\epsilon_{\vk}^{z}=-2t_{z} (\cos k_x + \cos k_y)$, 
$\mathcal{J}$ becomes positive, indicating that the major stacking is F. 
As a conclusion,  
for $\Lambda_{z}=0$ the stacking of the $d$FSD depends on the 
precise form of $\epsilon_{\vk}^{z}$.

\begin{figure}
\centerline{\includegraphics[width=0.35\textwidth]{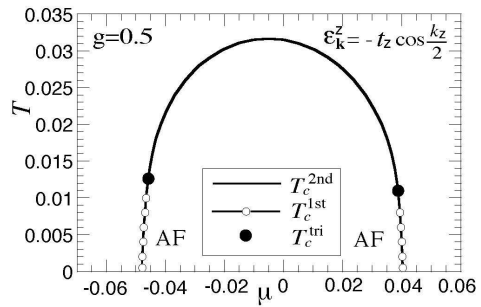}}
\caption{$\mu$-$T$ phase diagram in the infinite-layer model with 
$\epsilon_{\vk}^{z}=-t_{z}\cos \frac{k_{z}}{2}$ for $t_{z}=0.1$. 
The notation is the same as in Fig.~1.}
\label{3Dz0phase}
\end{figure}

To gain more insight into the physics of layer materials, we now 
analyze an infinite-layer model. 
This is given by the same 
Hamiltonian (\ref{model}) except that 
$t_{z}$ is replaced by $t_{z} \cos \frac{k_{z}}{2}$ and 
the $\vk$ summation is taken also for $k_{z}$; 
here we assume two layers per unit cell. 
The conclusions obtained from \eq{Landau-c3} still hold,
since the $k_{z}$ dependence appears only in $\epsilon_{\vk}^{z}$ 
in \eq{dzDOS} and just yields a factor of $\frac{1}{2}$ 
after the $k_{z}$ summation.  
AF stacking is thus stabilized in a major region of the phase diagram 
for $\Lambda_{z}\neq 0$. 
Unlike the situation in the bilayer model,  
bonding and antibonding bands now depend on $k_{z}$. 
In this case, even if the temperature scale of the $d$FSD 
becomes very small, the instability still occurs 
around $\mu=0$ \cite{misc3D}. 
The analysis of \eq{Landau-c3} then holds, yielding AF stacking 
[\fig{3Dz0phase}].

We have focused on a $c$ axis dispersion, which is certainly  
a relevant interlayer coupling in metallic systems. 
There may be other couplings such as 
Coulomb interaction $\sum_{i, j}V_{i j}^{z}n_{i}^{A}n_{j}^{B}$ 
and spin-spin interaction 
$\sum_{i,j}J_{i j}^{z}{\vS}_{i}^{A} \cdot {\vS}_{j}^{B}$.  
However, these interactions do not yield a direct coupling between 
$\eta^{A}$ and $\eta^{B}$. They may just shift 
the chemical potential and  renormalize 
the $c$ axis dispersion, which is indeed the case 
in the Hartree-Fock approximation,  
as long as the in-plane interaction $f_{\vk \vk'}$ is dominant. 

While our Hamiltonian (1) is simple, 
it is known that a forward scattering model works very well 
for Sr$_{3}$Ru$_{2}$O$_{7}$ \cite{kee05,doh07,yamase07bc,ho08} 
and that such a forward scattering interaction turns out to be inherent 
in the $t$-$J$ \cite{yamase00,miyanaga06,edegger06} and 
Hubbard \cite{metzner00,valenzuela01,hankevych02,neumayr03} models, 
minimal models for cuprate superconductors. 
In addition, although the symmetry 
is different from the $d$FSD, a forward scattering model 
is also known to work  for URu$_{2}$Si$_{2}$ \cite{varma06}.

The bilayer ruthenate Sr$_{3}$Ru$_{2}$O$_{7}$ is a promising 
material for the $d$FSD instability \cite{grigera04,borzi07}.  
LDA calculations \cite{liu08} show that  
the intrabilayer dispersion does not vanish at the saddle points 
$\vk=(\pi,0)$ and $(0,\pi)$, that is $\Lambda_{z}\neq 0$. 
Since the temperature scale of the $d$FSD is about 1 K and is much smaller 
than $\Lambda_{z}$ \cite{liu08}, we expect F $d$FSD stacking 
within each bilayer. 
On the other hand, there is a very weak interbilayer dispersion, 
which is expected to have the form factor 
$t_{z}'\cos\frac{k_x}{2} \cos\frac{k_y}{2}$ because the Ru sites shift 
by $(\frac{1}{2},\frac{1}{2})$ between adjacent bilayers. 
The interbilayer dispersion thus vanishes at 
$\vk=(\pi,0)$ and $(0,\pi)$, i.e. $\Lambda_{z}=0$. 
In this case, $\mathcal{J}$ is estimated as $\mathcal{J}\sim -1/\mu$ as 
calculated above. This suggests AF and F stacking  
between adjacent bilayers, namely the stacking $(++,--,++,--,\cdots)$ 
and $(++,++,++,++,\cdots)$ 
for $\mu \lesssim 0$ and $\mu \gtrsim 0$ \cite{misczeeman}, respectively.

High-temperature cuprate superconductors are also promising materials 
for the $d$FSD instability \cite{hinkov0407}. 
While competition with singlet pairing 
may suppress the instability, 
the $d$FSD is still important, yielding strong 
correlations \cite{yamase00,edegger06}, which may lead to 
a giant anisotropic response to a small external 
anisotropy, e.g. lattice anisotropy. 
Hence the stacking is determined by the 
external anisotropy and we expect 
F stacking for Y-based cuprates \cite{yamase06} and 
AF stacking in La-based cuprates in the low temperature 
tetragonal structure \cite{yamase00}.

We have studied a generic tendency for the $d$FSD stacking 
in bilayer and infinite-layer models. 
We have found that the stacking  
does not depend on the details of interlayer coupling 
as long as the $c$ axis dispersion is finite at the saddle points 
of the in-plane band dispersion, namely as long as $\Lambda_{z}\neq 0$. 
For bilayer systems  we have found 
F stacking only when the temperature scale of the 
$d$FSD is much smaller than $\Lambda_{z}$ [\fig{z0phase}(d)], 
and AF stacking otherwise [Figs.~\ref{z0phase}(a)-(c)]. 
For infinite-layer systems instead the major stacking pattern is 
always AF [\fig{3Dz0phase}].

The $d$FSD instablity is a generic tendency in correlated electron systems 
as found in different theoretical models \cite{yamase00,metzner00,valenzuela01,quintanilla06}.   
The $d$FSD is thus expected for various materials 
except if other instabilities prevail over the $d$FSD instability.  
Since the condition $\Lambda_{z}\neq 0$ is fulfilled for 
most materials unless there is a special symmetry reason, 
typical materials with a high critical temperature for the $d$FSD instability 
may have AF stacking. 
While spontaneous symmetry  breaking is by definition 
recognized by symmetry breaking of the system, 
the $d$FSD provides a conceptually interesting 
property that spontaneous symmetry breaking can be self-masked in realistic materials.

Given the fact that antiferromagnetism was recognized 
about twenty centuries later than ferromagnetism \cite{neel32,shull49}, 
it is not difficult to believe that the $d$FSD instability is 
hidden in various layer materials. 
Because of a coupling to the lattice, 
the easiest step to find AF $d$FSD materials is to test 
a twofold lattice modulation along the $c$ axis.

The author is indebted to W. Metzner for supporting the 
present work, and to O. K. Andersen and G.-Q. Liu 
for extracting the tight binding parameters for 
Sr$_{3}$Ru$_{2}$O$_{7}$. He also acknowledges 
G. Khaliullin, G. Sangiovanni, and R. Zeyher 
for very useful discussions.



\bibliography{main.bib}

\begin{thebibliography}{29}
\bibitem[{mis({\natexlab{a}})}]{misc-kzdispersion}
\bibinfo{note}{One could consider the $c$ axis dispersion
  $\epsilon_{\vk}^{z}=-t_{z}-2t_{z}^{'} (\cos k_x + \cos k_y)$ with $|t_{z}| <
  |t_{z}^{'}|$, for which we cannot infer AF stacking despite $\Lambda_{z}\neq
  0$, since the second term favors F stacking, which competes with AF stacking
  driven by the first term. But the condition, $|t_{z}|< |t_{z}^{'}|$, is
  unrealistic. Our conclusion holds for a physical parameter region such as
  $|t_{z}| > |t_{z}^{'}|$.}
\bibitem[{mis({\natexlab{b}})}]{misc3D}
\bibinfo{note}{While the $d$FSD instability happens for an arbitrary $g$ in the
  bilayer model as in the case of the single-layer model
  (Ref.~\onlinecite{yamase05}), there is in general a threshold coupling
  constant $g_{\rm th}$ in the infinite-layer model: $g_{\rm th} \sim 0.4$ and
  $0.2$ for $\epsilon_{\vk}^{z}=-t_{z}\cos\frac{k_{z}}{2}$ and
  $-4t_{z}\cos\frac{k_{x}}{2}\cos\frac{k_{y}}{2}\cos\frac{k_{z}}{2}$,
  respectively, for the present choice of the parameters. Although the value of
  $g$ in Fig.~3 is rather close to $g_{\rm th}$, $T_c$ is still comparable to
  that in the single-layer case.}
\bibitem[{mis({\natexlab{c}})}]{misczeeman}
\bibinfo{note}{The chemical potential should be read as a magnetic field for
  Sr$_{3}$Ru$_{2}$O$_{7}$; the field effect is studied in
  Ref.~\onlinecite{yamase07bc}.}
\end{thebibliography}

\end{document}